\documentclass[times]{cpeauth}
\usepackage{moreverb}
\usepackage{color}
\usepackage[listings,skins]{tcolorbox}
\tcbuselibrary{fitting}
\usepackage{lineno}
\usepackage{relsize}

\newtcolorbox[auto counter, number within=subsection]{pabox}[2][]{%
  title={\footnotesize Example~\thetcbcounter: #2}, #1}

\usepackage[colorlinks,bookmarksopen,bookmarksnumbered,citecolor=red,urlcolor=red]{hyperref}
\newcommand\BibTeX{{\rmfamily B\kern-.05em \textsc{i\kern-.025em b}\kern-.08em
T\kern-.1667em\lower.7ex\hbox{E}\kern-.125emX}}
\def\volumeyear{2015}

\newcommand{\lit}[1]{{\relscale{0.9} \tt #1}}

\begin{document}

\runningheads{P.~Huck \emph{et~al.}}{MPContribs-driven User Applications}
\title{User Applications Driven by the Community Contribution Framework
  MPContribs in the Materials Project}
\author{P.~Huck\corrauth, D.~Gunter, S.~Cholia, D.~Winston, A.~T.~N'Diaye\ and
  K.~Persson\affil{1}}
\address{\affilnum{1}Lawrence Berkeley National Laboratory, Berkeley,
  California 94720} 
\corraddr{Lawrence Berkeley National Laboratory, 1 Cyclotron Road, M/S 33R0146,
  Berkeley, CA 94720. E-mail: phuck@lbl.gov}

\begin{abstract}
  This work discusses how the MPContribs framework in the Materials Project
  (MP) allows user-contributed data to be shown and analyzed alongside the
  core MP database. The Materials Project is a searchable
  database of electronic structure properties of over 65,000 bulk solid
  materials that is accessible through a web-based science-gateway. We describe
  the motivation for enabling user contributions to the materials data and
  present the framework's features and challenges in the context of two real
  applications. These use-cases illustrate how scientific collaborations can
  build applications with their own ``user-contributed'' data using MPContribs.
  The Nanoporous Materials Explorer application provides a unique
  search interface to a novel dataset of hundreds of thousands of materials,
  each with tables of user-contributed values related to material adsorption
  and density at varying temperature and pressure. The Unified Theoretical and
  Experimental x-ray Spectroscopy application discusses a full workflow for the
  association, dissemination and combined analyses of experimental data from the Advanced Light Source
  with MP's theoretical core data, using MPContribs tools for data formatting,
  management and exploration. The capabilities being developed for these
  collaborations are serving as the model for how new materials data can be
  incorporated into the Materials Project website with minimal staff overhead
  while giving powerful tools for data search and display to the user community.
\end{abstract}

\keywords{Science Gateways; Databases; User Contributed Data; Materials Science}

\maketitle

\section{Introduction}

In recent years, an exciting trend has begun to emerge - the integration of
computational materials science with information technology (e.g., web-based
dissemination, databases, data-mining) to go beyond the confines of any single
research group. This development has expanded access to computed materials
datasets to new communities and spurred new collaborative approaches for
materials discovery. The next step is to leverage open-source development and
interactive web-based technologies to enable user contributions back to the
community by reporting problems, coding new types of analyses and apps, and
suggesting the next set of high potential materials for computation. All the
pieces are ready to change the paradigm by which materials are designed.

Materials Project (MP)~\cite{Jain2013} is a component of the Materials Genome
Initiative~\cite{WHMGI2014} that leverages the power of high-throughput
computation and best practices from the information age to create an open,
collaborative, and data-rich ecosystem for accelerated materials design.
Started in October of 2011 as a joint collaboration between the Massachusetts
Institute of Technology and Lawrence Berkeley National Laboratory, the
Materials Project today has partners in more than ten institutions worldwide.
The Materials Project web site currently receives several hundred unique page
views a day and has registered more than 13,000 users in academia, government,
and industry.

MP uses high-performance computing (HPC) to determine structural,
thermodynamic, electronic, and mechanical properties of over 65,000 inorganic
compounds by means of high-throughput ab-initio calculations. The calculation
results and analysis tools are disseminated to the public via modern web and
application interfaces.  However, the materials science research community has
a continually increasing supply of experimental and theoretical material
properties that are either not yet calculated by MP or outside the scope of
MP's data generation efforts. It becomes increasingly important for MP and
similar scientific platforms to also enable community-driven submissions, which
would extend the scope of the possible applications and improve the
integrity/quality of the provided datasets, hence enhancing its value to the
user community.

In this paper, we describe how the MP web portal manages user contributions
and dissemination of data. In particular, we discuss how our
framework for user contributed data, called
MPContribs~\cite{mpcontribs_escience}, is being extended to handle two
major new use-cases: (1) simulation results for large numbers of new objects
(i.e., crystals) with very different properties of interest and (2)
experimental results that augment the simulation data for existing objects.
Both these use-cases have general aspects that recur in other scientific
projects, but here we describe the challenges and solutions in the context of
two specific projects: for (1), incorporation of millions of nanoporous
materials from the Nanoporous Materials Genome Project (NMGC) and for (2),
analysis of high throughput spectroscopy data from the Advanced Light Source
(ALS) at LBNL. Each use-case illustrates different aspects of the flexibility
and performance of the MPContribs framework.

Most of the tools described here (including MPContribs and MPContribsUsers) are
released as open source software and are available on the MP Github repository
at https://github.com/materialsproject for download.

\section{Materials Project Portal}

The Materials Project portal consists of a web portal that interfaces with a
large database of materials, stored in a MongoDB NoSQL database
~\cite{mongodb}. We chose MongoDB for its rich and flexible data model, which
allows us the flexibility to evolve our data structures and schemas as we add
new materials and materials properties from different sub-domains of the
field~\cite{Gunter_communityaccessible}.

\begin{figure}[htbp]
\centering
\includegraphics[width=\textwidth]{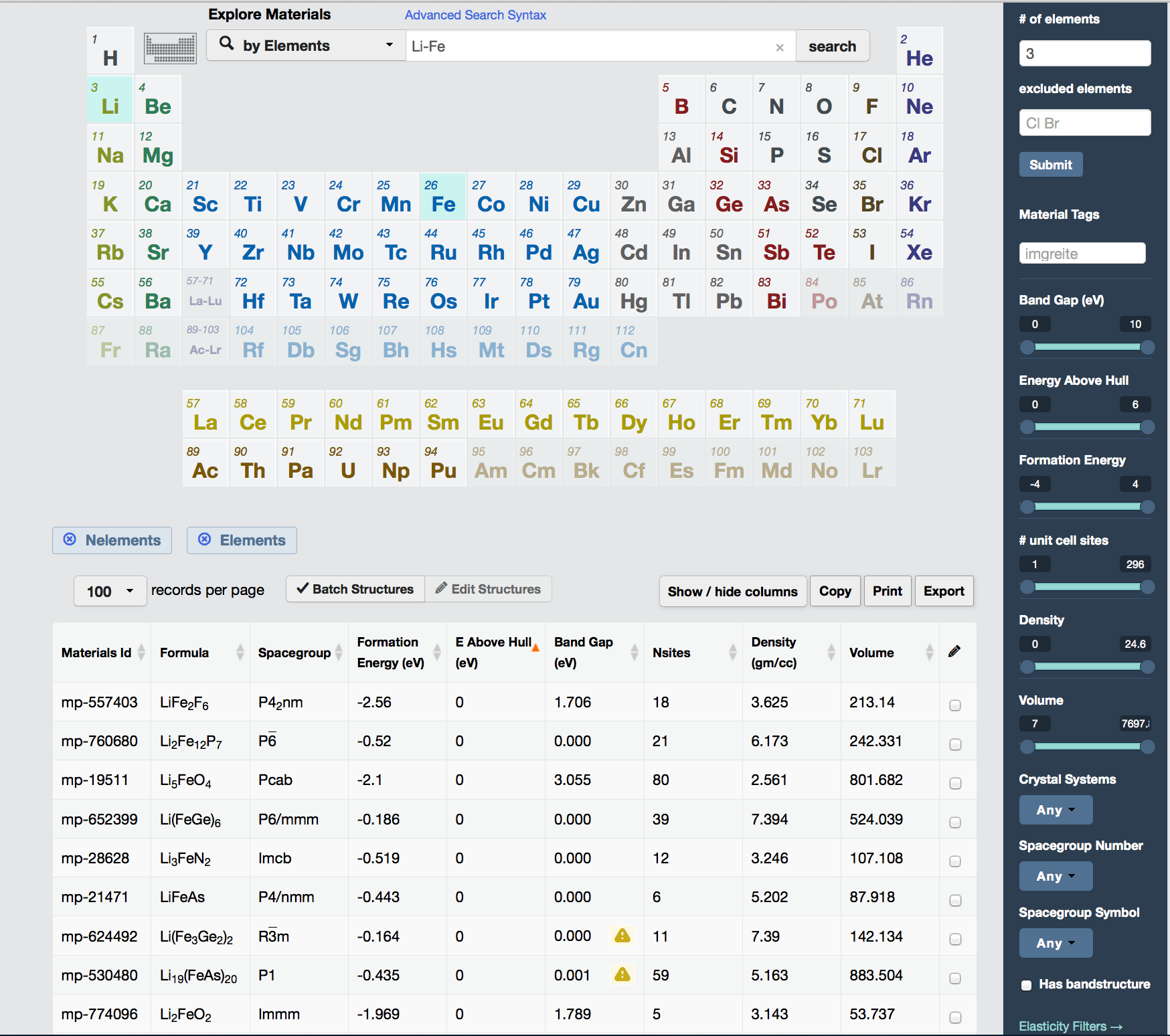}
\caption{The Materials Explorer application provides an intuitive interface to
  search for materials in the Materials Project (MP) database. Users select
  elements from a periodic table and refine their queries to look for
  properties within specific ranges.}
\label{fig_matexplorer}
\end{figure}

The portal is built using the Python Django web framework.
It is hosted at the National Energy Research Scientific
Computing (NERSC) Center of LBNL. The data are
generated through a series of high-throughput computational runs at NERSC by
performing density-functional theory calculations on a large space of possible
materials. The results of these calculations are aggregated and indexed in
MongoDB, which can then be queried using the web portal or through a REST API
~\cite{ong2015materials}. Figure~\ref{fig_matexplorer} shows the search
interface to the MP data - users enter search parameters such as the elements
they are interested in, along with specific materials properties and the portal
returns a set of matching results. Users can then drill down and look at the
data for a specific material that they might be interested in.

The MP portal supports a series of applications that lets users explore
materials data. Along with the aforementioned Materials Explorer, there is a
Lithium Battery Explorer which adds application-specific search criteria for
lithium-ion battery electrode materials.  Additional web applications allow
users to interactively analyze the dataset. For instance, the Phase Diagram App
constructs low-temperature phase diagrams for any chemical system. The Reaction
Calculator balances reactions and computes reaction energies between sets of
compounds in the database. Other applications include the Pourbaix Diagram
viewer, the Crystal Toolkit, the Structure Predictor and the Nanoporous
Materials application (which we discuss in detail in Section~\ref{nanoporous}).

MP also supports a REST API, which allows more sophisticated users to
generate custom queries, perform bulk downloads of data and interface with the
data in a programmatic fashion.

A common concern in scientific collaboration is the sharing of data among
privileged collaborators prior to its wider dissemination after
validation/publication. MP supports such ``sandboxed'' data exploration through
its REST API as well as with additional web applications such as the Molecule
Explorer for battery electrolyte materials~\cite{qu_eg_2015}, which is not
currently available to all MP users. However, another pattern of data sharing,
whereby contributors wish not to maintain a privileged sandbox but rather
intend to make data publicly available -- perhaps after a short incubation
period for inspection -- is the aim of the MPContribs framework.

\section{MPContribs Framework}
\label{mpcontribs_overview}

The current pipeline consists of calculations run directly by MP staff, at
NERSC and other computing centers, the results of which are fed directly into
the core database using a set of automated builder scripts. MP represents a
growing community of materials scientists, and one of the key features towards
making this project sustainable is the ability to integrate user contributed
data directly into MP. The longer term vision of MP is to be a hub for
datasets that come from several different contributors, and are integrated
under a common platform.

Towards this end, we have developed a framework called MPContribs that
enables users to add their own data to MP, and use the underlying tools to
interact with and reference that data.

\begin{figure}[t]
\centering
\includegraphics[width=\textwidth]{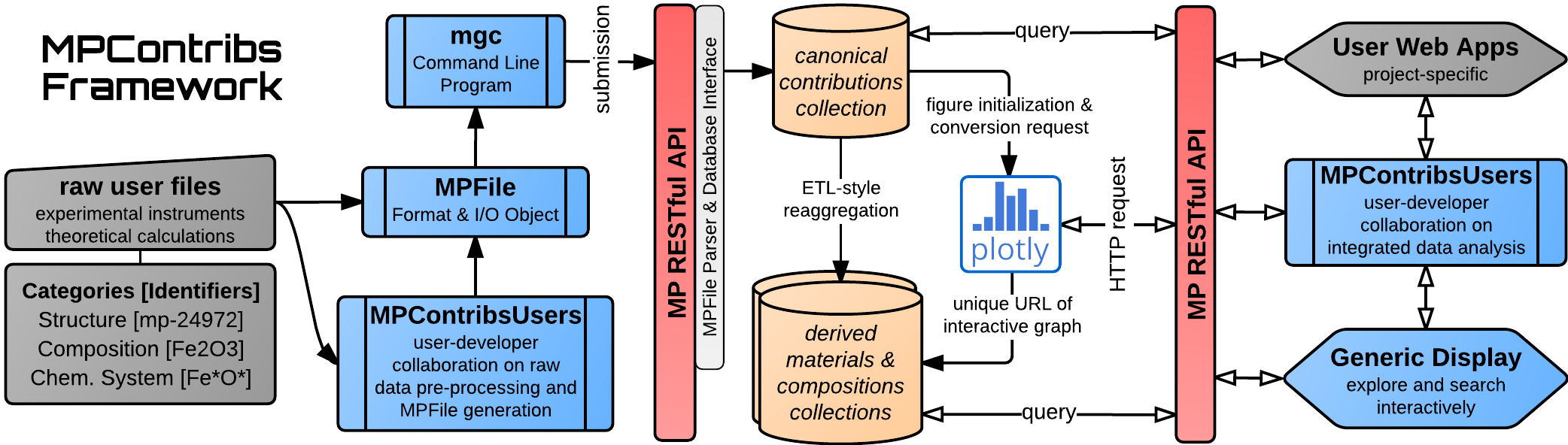}
\caption{Processing workflow of \texttt{MPContribs} framework. The
  pre-submission stages, with category identification and \texttt{MPFile}
  generation, are followed by internal parser and builder phases and the
  dissemination to the interactive portal. Contribution submission as well as
  queries to the databases (orange) and Plotly requests operate through the
  MP's REST API (red). Blue boxes and the processing behind the REST API "wall"
  are provided by the \texttt{MPContribs} framework. Grey boxes denote tasks and
  tools under the user's responsibility. For more details, see text.}
\label{fig_flow_chart}
\end{figure}

We briefly summarize the MPContribs framework here; details on
the format and tools provided by the MPContribs framework are
described in~\cite{mpcontribs_escience} and available
online~\cite{mpcontribs_url,mpcontribsusers}. The three major features of the
framework are (1) a text format and associated parsing tools that can flexibly
incorporate user-defined data, (2) a RESTful API that can receive and update
records in this format to a back-end database, and (3) a display framework for
showing the data in the context of MP core databases. The workflow connecting
these components is shown in Figure~\ref{fig_flow_chart}. First, the raw user
output is identified and gathered. Examples of such data include the nanoporous
materials and x-ray absorption (XAS) spectra discussed below, as well as
diffusivities computed for various temperatures and solutes, band gaps
calculated under alternative conditions, and electronic structure calculations
for photovoltaics. Next, the raw data are converted into a concise summary using
the MPFile syntax. The result of this phase is submitted, by the user,
via the command line program or web interface, both of which operate through
the RESTful MP API. The received data are parsed, and the resulting objects are
correlated with existing materials data via suitable identifiers. The results
are inserted into the core database. Finally, the data are disseminated through
the API---and a graphical UI with interactive graphs, tables and tree
structures---for exploration and analysis.

As a collaborative extension to the MPContribs core, we created
MPContribsUsers~\cite{mpcontribsusers}, a codebase hosted on Github. It
contains reusable routines serving a dual purpose: \textit{i}) pre-submission
processing to help with the conversion of raw data into MPFiles, and
\textit{ii}) analyzing and post-submission processing of contributed data in an
integrated manner with MP core data. By hosting this dual-purpose software on
Github, the standard open-source contribution workflow\footnote{fork repo, make
  edits \& issue a pull request for merging w/ master repo} can be employed
easily with our collaborators which sets up MPContribsUsers as a designated
central area separated from the MPContribs core. It consequently allows for the
collaborative involvement of the framework developer while simultaneously
establishing an archive of example files and scripts to support future
contributors. It also enables coordinating between users and reducing
duplication of code that deals with similar or identical formats in different
domains of materials science.

The Nanoporous Materials Explorer (Section~\ref{nanoporous}) and the X-ray
Spectroscopy Application (Section~\ref{xrayspecs}) sections illustrate how this
is used in practice.

\section{Nanoporous Materials Explorer}
\label{nanoporous}

The Nanoporous Materials Genome Center (NMGC)~\cite{NMGC} discovers and
explores microporous and mesoporous materials, including metal-organic
frameworks (MOFs), zeolites, and porous polymer networks (PPNs). These
materials are important for many applications, including separation media and
catalysts in many energy-relevant processes. The NMGC provides a repository of
experimental and predicted structures and associated properties for the rapidly
growing scientific communities that are interested in using these materials in
energy-relevant technologies.

The Materials Project has integrated porous material data from the NMGC with
its software infrastructure. These data are accessible through the production
MP website via a \textit{porous materials ``app''}~\cite{nmgc-app} that allows
for interactive browsing of the multidimensional space of porous materials
properties. This graphical search interface allows the user to plot the values
of pairs of properties, pan and zoom on the graph, and then list all materials
in a region of interest, and select from that list for a detailed per-material
view. This interactive view of potentially millions of materials requires its
own database of discretized values in each dimension on the website back-end.

The current database and porous materials app were created by custom loading
scripts. The indexing of the database for efficient graphical search is
performed by a post-processing script in the Materials Project database
``builder'' framework, which is used for post-processing of electronic
structure and other properties. The rest of this section will describe how we
are moving from this custom import of NMGC data to using the MPContribs
framework to regard all the porous material properties as ``contributed data''
on the base crystal structure, and to automate the post-processing through the
MPContribs maintenance and analysis tools. This allows updates to the data to
be performed independently by NMGC collaborators
through the increasingly sophisticated interfaces provided by the MPContribs
framework. In the longer view, it also will allow multiple types of data to be
added to porous materials in the same way they are added to MP's ``core''
materials. From a devops perspective, this lessens the maintenance burden of
keeping up two different databases.

The NMGC data provides an example of how a well-structured custom data format
can easily be adapted to MPContribs. In this case, the input data has a crystal
structure (giving the position of each atom in the unit cell) that corresponds
exactly to core MP crystal structures, and then a set of properties relevant to
its application such as the \emph{heats of adsorption}, \emph{crystal density},
\emph{Henry coefficient}, and \emph{isotherms} for the loading of the material
for a given gas molecule at varying pressure. All of these data types are
easily represented as multi-column tables with attached metadata, and therefore
we are able to perform the conversion from our existing data to MPContribs
files automatically. The identifiers for the materials will be of the form
\lit{<prefix>-<nativeId>}, where the prefix is specific to porous materials
(e.g., ``por'') and the \lit{nativeId} is the identifier that is consistently
used by the software that generates the porous materials properties. This
allows updates to use these native material identifiers.

An important requirement of the MPContribs integration is that the porous
materials remain separate from the core set of materials for the purposes of
searching and display, that is, they can only be accessed through the porous
materials app. In the short term, we are doing this via some ad-hoc routing
using our pre-assigned material identifiers; the NMGC materials will remain in
their separate database. But the long-term goal is to avoid special
databases for different sets of materials properties, so we are looking into
using the existing Materials Project ``sandboxing'' capability, which can
restrict certain materials properties to specific groups. In this
case a pseudo-user (i.e. porous materials app), can be used to
protect the porous materials data even if they are in the core database.
Another potential advantage of this deeper integration is that the core
Materials Project explorer app could also, if desired, include porous materials
in searches by selected users, by simply changing group memberships.

Another important aspect of the NMGC data that must be preserved is the
post-processed discretization of the data values that allows for efficient
retrieval and zooming in the search interface. This is easily created with
existing scripts for the initial data load, but for updates of the data the
scripts will become part of MPContribsUsers to be co-maintained with the
framework developer using pull requests (see
Section~\ref{mpcontribs_overview}). Once the pull request is merged, the
framework will
call the user's post-processing methods automatically on a set of materials for
which he has sufficient permissions. This is achieved by a REST API endpoint
exposing routines in the MPContribs core that allow the user to run limited
post-processing algorithms based on abstract classes in the core library.

\begin{figure}[p]
\centering
\includegraphics[height=0.9\textheight]{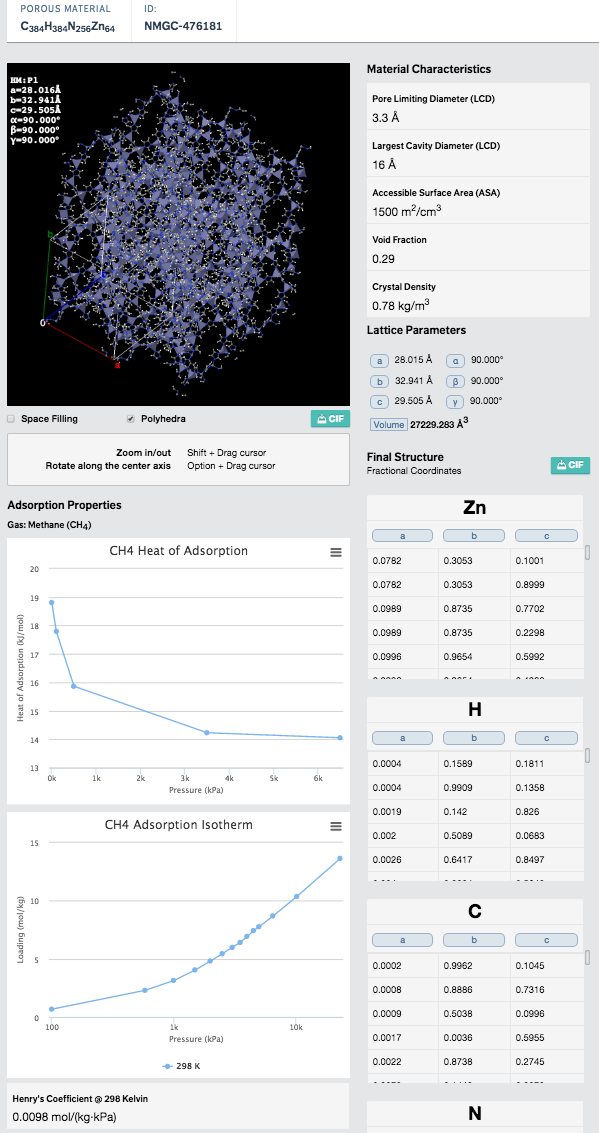}
\caption{Current detail view for a porous material. The view template was
  custom-built to display porous material data. Note that the data on display,
  beyond those pertaining to the crystal structure, are a set of key-value
  pairs (tree-like data) and simple 2D plots (derived from tabular data). Such
  a display may be prototyped by a contributor using the generic front-end of
  MPContribs prior to optional customization.  }
\label{fig_porous_detail}
\end{figure}

A major advantage of using MPContribs for this data is the ability to use the
MPContribs generic display interface, in lieu of the current custom display,
for the details of each material. Figure~\ref{fig_porous_detail} shows the
current detail view for a porous material. The view template was custom-built
to display porous material data, but note that the data on display, beyond
those pertaining to the crystal structure, are a set of key-value pairs
(tree-like data) and simple plots (derived from tabular data). Such a display
may be prototyped by a contributor using the generic front-end of MPContribs,
making customization optional. A screenshot of a prototype \textit{Contribution
  Details Page} is shown in \cite{mpcontribs_escience} featuring hierarchical
and tabular data along with interactive and customizable graphs as well as
search capabilities.

An important challenge introduced by these materials is bulk operations. The
porous materials are generated by a method very different from the core MP
database, which is based on a database (the ICSD~\cite{icsd}) culled from many
years of theory and experiment. The porous materials can be generated and
analyzed very quickly, so that hundreds of thousands of new materials, or sets
of new properties on existing materials, can be created. This poses a
challenge to the current MPContribs interfaces, which operate on a single
material at a time. We are working on a bulk upload and bulk modify interface
that will remove the round-trips for large numbers of materials.

\section{Unified Theoretical \& Experimental x-ray Spectroscopy Application}
\label{xrayspecs}

As discussed in Section~\ref{mpcontribs_overview}, the MPContribsUsers
repository collects code (\textit{i}) for data generation to transform output
files from common simulation software or from experimental instruments into the
MPFile format, and (\textit{ii}) for integrated data analyses to unify MP core
data and MPFile contributions.

An example of codes for data generation that would naturally belong in this
repository are domain-specific multi-level directory parsers to compile the
extracted and derived data into MPFiles. With directories containing electronic
calculations based on crystal structures, for instance, the parser can use
\textit{pymatgen}'s~\cite{Ong2013} powerful \texttt{Borg} module for
high-throughput data assimilation as well as its structure matcher code to
match the input structures with existing MP structures of the core dataset.
This allows automatic assignment of MP materials identifiers from
general-purpose crystal structure definitions and its accompanying
calculations.

In the remainder of this section, we discuss the contribution and analysis of
measured x-ray absorption spectra as an example which uses both aspects of
MPContribsUsers - data generation and data analysis. The result is a
MPFile-based and pipelined dataflow as well as analysis workflow established in
collaboration with our experimental colleagues.\\

Soft x-ray absorption spectroscopy (XAS) is a characterization technique which
probes the electronic structure of materials. It yields element-, valence-, and
lattice site-specific information.
At beamline 6.3.1 at the Advanced Light Source (ALS), XAS
is being implemented as a high throughput technique~\cite{Idzerda2015} yielding
datasets with a few hundred spectra per measurement.

Each measurement produces a text file containing data which corresponds to
several spectra taken on a certain material and with
instrument specific technical information (meta-data). This set of
associated instrumental output files is combined with their shared meta-data
from the experimentalist's lab book using the MPFile format, into a single output file.
This output file also serves as a configuration/input file to apply a sequence
of automated pre-processing routines to clean and refine the raw data for a
large selection of measured compositions. These pre-processing routines embed
their results into the MPFile using composition, e.g. $Ni_{20}Fe_{80}Pt_{10}$,
as universal cross-domain identifier. The embedded data can be
tabulated datasets, such as XAS spectra or sets of numbers, such
as lattice parameters or magnetic moments. Example \ref{myautocounter} shows an
excerpt of a MPFile containing meta-data, processing instructions and embedded
spectral data measured on Permalloys to study their change in magnetic
properties with increasing Platinum doping.

\begin{figure}[t]
\centering
\includegraphics[width=\textwidth]{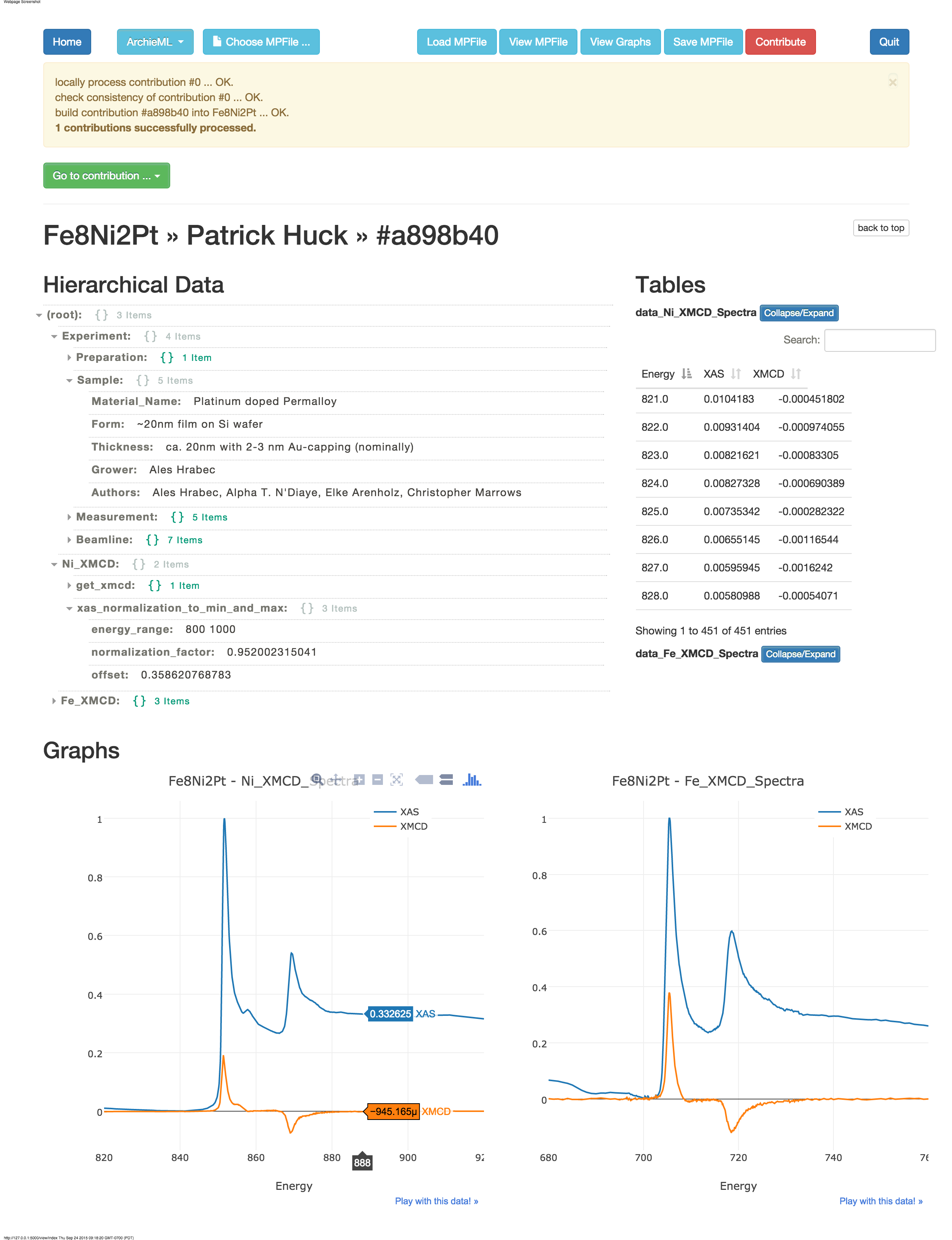}
\caption{Screenshot of \texttt{MPContribs}' offline MPFileViewer web app
  showing the contribution with composition Ni20Fe80Pt10.}
\label{fig_viewer}
\end{figure}

Before official submission of the final MPFile to MP and dissemination of the
contributed data to its users, the contained data can be inspected locally
using a software tool called \textit{MPFileViewer}. This tool is a Flask web
application that simulates a submission using the same internal methods on the
local machine.  The ability to debug results locally in a realistic environment
is integral in the preparation of data for its submission to a well-maintained
community database.  A screenshot of the XAS data rendered by the MPFileViewer
is depicted in Figure \ref{fig_viewer}. The contribution can be explored via
collapsable tree and table structures as well as interactive graphs. The
submittors can subsequently use the MP web portal with its dedicated
\textit{Contribution Detail Pages} for further analysis and to confirm links to
existing projects or update the dataset after additional local analysis steps.

\vspace{1em}
\scalebox{0.9}{
\begin{pabox}[label={myautocounter}]{Sample \texttt{MPFile} content for
measured XAS data on composition Ni20Fe80Pt10. Line numbers and bold section
headers were added for presentation purposes. The representation of the MPFile
in the \texttt{MPFileViewer} is shown in Figure~\ref{fig_viewer}. The content
of the \texttt{GENERAL} section is embedded into all other root-level sections
to enable common/shared meta-data. Lines in italic font are added by the
pre-processing scripts via \texttt{MPContribs} tools based on a skeleton input
\texttt{MPFile} and instrumental output files (see Section~\ref{xrayspecs}).}
\tcbfontsize{0.85}%
\setlength\linenumbersep{5pt}%
\internallinenumbers%
\textbf{\texttt{>>>} GENERAL}\\
\textbf{\texttt{>>>>} Experiment}\\
\textbf{\texttt{>>>>>} Preparation}\\
Description: Sputter deposition\\
\textbf{\texttt{>>>>>} Sample}\\
Material Name: Platinim doped Permalloy\\
Form: ~20nm film on Si wafer\\
Thickness: ca. 20nm with 2-3 nm Au-capping (nominally)\\
Grower: Ales Hrabec\\
Authors: Ales Hrabec, Alpha T. N'Diaye, Elke Arenholz, Christopher Marrows\\
\textbf{\texttt{>>>>>} Measurement}\\
Detection: total electron yield\\
Temperature: RT\\
Orientation: 30° grazing incidence\\
Date: 2015-06-24\\
Measured by: Alpha T. N'Diaye\\
\textbf{\texttt{>>>>>>} Beamline}\\
Beamline: ALS-6.3.1\\
Method: Soft x-ray XAS and XMCD\\
Polarization: circular, positive (ca. 60\%)\\
Magnet Field: 0.8T switching point by point, parallel to x-ray beam\\
Count Time: 1s\\
Delay Time: 0.5s\\
\textbf{\texttt{>>>>>>>} Monochromator}\\
Exit Slit: 20µm\\
Grating: 600l/mm\\

\textit{\textbf{\texttt{>>>} Ni20Fe80Pt10}}\\
\textbf{\texttt{>>>>} Ni XMCD}\\
\textbf{\texttt{>>>>>} get xmcd}\\
energy range: 800 1000\\
\textbf{\texttt{>>>>>} xas normalization to min and max}\\
energy range: 800 1000\\
\textit{normalization factor: 0.952002315041}\\
\textit{offset: 0.358620768783}\\
\textit{\textbf{\texttt{>>>>} Ni XMCD Spectra}}\\
\textit{Energy,XAS,XMCD}\\
\textit{820,0.0104944,-0.00140602}\\
\textit{821,0.0104183,-0.000451802}\\
\textit{822,0.00931404,-0.000974055}\\
\textit{...}\\
\textbf{\texttt{>>>>} Fe XMCD}\\
\textbf{\texttt{>>>>>} get xmcd}\\
energy range: 600 800\\
\textbf{\texttt{>>>>>} scaling preedge to 1}\\
preedge range: 690 700\\
\textit{xas- factor: 0.348231766387}\\
\textit{xas+ factor: 0.349333591384}\\
\textbf{\texttt{>>>>>} xas normalization to min and max}\\
energy range: 600 800\\
\textit{normalization factor: 1.00964185927}\\
\textit{offset: 0.984095999176}\\
\textit{\textbf{\texttt{>>>>} Fe XMCD Spectra}}\\
Energy,XAS,XMCD\\
680,0.0670848,0.000905727\\
681,0.0659347,-0.00085033\\
682,0.0631599,-8.87504e-05\\
...
\end{pabox}
}

We require that all the processing routines used to prepare the input (in this
case, the experimental x-ray data) be part of the MPContribs libraries. This
way, other user contributions can link to each-other without losing provenance
or reproducibility of the work. The data on the MP servers can be embedded in
the analysis iterations seamlessly, eventually leading to publishable data
which is already linked to relevant work in the field. The division of
responsibility for software development allows independent groups, such as
submitters with a focus on x-ray spectroscopy, to add their own routines, while
keeping broader tools such as the MPFile API and the associated offline
inspection and manipulation tools within the purview of the MP team.

The code in MPContribsUsers can be extended to perform advanced functions.  For
example, one goal of the x-ray spectroscopy application is to compare the X-ray
magnetic circular dichroism (XMCD) signals resulting from the XAS spectra with
the corresponding theoretical phase diagrams calculated by MP. To achieve such
a unified analysis/view of experimental and theoretical data, the post-submission
workflow looks something like this: Gather contributed
XAS/XMCD spectra and magnetic moments for related compositions from the MP
database; Generate an overview PDF including a heatmap of magnetic moments;
Pull the according MP phase diagram; Produce an instructive comparison
graph. This can be implemented collaboratively in MPContribsUsers, using
either scripts or, more attractively, via a customized dynamic web app.

The depth of information in the Materials Project database will enable exciting
future scientific directions for this contributed data. We expect that
experimental contributors will be able to study their submitted absorption
spectra with respect to FEFF~\cite{feff} calculations to relate the measured
compositions to existing crystal structures.

\section{Contribution Process}

The purpose of MPContribsUsers repository is to collect pre- and
post-submission codes primarily running on the user's local infrastructure.
Since the code would live in MPContribsUsers it is up to the core developer
to make use of it (pull-based approach). As such, any new code would
need to be vetted if it is merged back into the core. Customized web apps
should also be deployed on the user's infrastructure. MPContribs, however,
would provide the tools to easily start up web apps that are already integrated
with core and contributed MP data.

Currently, we limit contributions to known collorators who are trusted domain
experts to produce curated, high-quality datasets. At this stage, our goal
is to provide tools for efficient manual data inspection. For instance,
we have a staging area for data contributions (MPFileViewer), which allows for
human verification before release on the live web site. We have also integrated
the plotting tool "Plotly" with our service, to visualize results and reveal
outliers or incorrect data.

In the near future, we plan to move to automated self-verification
of the data. This includes a data access API, along with instructions on how users
can employ the API to develop unit testing and continuous integration (CI) frameworks.
This type of CI functionality will greatly improve the robustness and reliability of
user-submitted datasets. We are also investigating other possiblities, like
user ratings for datasets. This would enable mechanisms
like flagging, removal or black-listing of consistently low-quality datasets/users.

Note, that the materials data format used by MP already has fields for citations and references,
thus providing provenance and source tracking for contributed data.

\section{Related Work}

There are a number of community repositories in Materials Science and related
disciplines. Traditionally, these repositories have been limited in scope to
crystal structures of measured compounds (acting analogously to the Protein
Data Bank~\cite{Berman2000} for biology), and include efforts such as the
Inorganic Crystal Structure Database (ICSD)~\cite{Belsky2002} 
and Pauling File~\cite{pauling}. More recent efforts, such as
AFLOWlib~\cite{aflowlib} and
others~\cite{cccbdb,comp-es,ortiz2009data,cepdb,oqmd}, typically include
additional property data (generally computed) but do not currently have a
contribution framework.  Some repositories, such as NoMaD~\cite{nomad},
iochem-bd~\cite{moreno2015managing}, and the Materials Data Facility
(MDF)~\cite{mdf}, provide a shared repository of data, with potentially large
experimental and structural data items, but many of these are in the early
stages of development and without integrated data analysis capabilities over a
core dataset. Our approach, by contrast, focuses on relatively small data
contributions that can be joined with the carefully curated core
  dataset, and extended with an integrated set of data analysis tools to
provide a materials design environment.

A number of projects in other sciences, particularly life sciences, share the
goals of community-enabling data contributions and analysis in the context of
reference datasets. The iPlant Collaborative~\cite{iplant} combines a "Data
Store" with a ``Discovery Environment'' with the goal of connecting public and
private datasets with analysis capabilities; the analysis capabilities include
workflows running on remote clusters and private Virtual Machines. The
DOE's Systems Biology Knowledgebase (KBase)~\cite{kbase}
provides a unified and interactive analysis environment built on the IPython
notebook~\cite{perez2007ipython} and a back-end store that can store users'
analyses together with both their data and standard reference microbial data.
The large amount of funding
spent by federal agencies (KBase and IPlant each receive over \$10M/year) on these large
cross-cutting projects points to the importance of the capability. Our work is
scoped more narrowly (e.g., it does not require an overarching data model or remote execution
engines), and is simplified by the use
of a core identifer -- the \lit{material id} -- that represents
a common object, a crystal structure, to organize contributions. A genome, the logical
analogue of a crystal in biology, is orders of
magnitude more complex and variable and cannot be used easily as a central identifier.

Efforts such as ChemSpider~\cite{chemspider} emphasize aggregation of data from
many individual sources rather than extension of ``core'' reference data.
However, the success of such efforts depends on widely-used standards for
identifiers~\cite{Heller2013} and does not allow users to contribute and
maintain custom datasets. Thus, our framework occupies a unique
position in the materials data sharing space.

\section{Summary \& Outlook}\label{sec_sumout}

With over 13000 (and growing) users, the Materials Project has been very
successful in serving as a hub for the materials science community. As MP moves
forward, user contributed data will play an important role in expanding the
project towards serving a larger community. By providing a standard framework in
the form of MPContribs and a standard file structure in the form of MPFile, we
are providing a common interface for users to add data to the project. We
believe that this provides the basis for other projects that wish to add data to
MP and to build custom web applications on top of this data. As illustrated by
the use cases above, this is already being used to actively  develop new web
applications in the MP portal to serve up custom data.

Note that the contributions format and the standard material database objects already
keep track of provenance information to track details like data sources and collaborators.
Furthermore, our choice of a document oriented database like MongoDB makes it very
easy to add new fields to an object to track additional provenance metadata.
We expect provenance tracking to become an important requirement for future efforts,
and are well positioned to handle this.

In conclusion, MP has delivered significant value to the materials science
community by creating a general purpose web gateway for querying and accesing various
materials datasets. As newer datasets get incorporated into the MP database, there is
a need to be able to build more specialized applications that target specific
classes of materials. With our work we are creating the foundation for a common platform
that can be used to host multiple applications and datasets for different user communities.

\acks
We especially thank Anubhav Jain for guidance and help with the manuscript.\\
This work was intellectually led by the Department of Energy's Basic Energy
Sciences program - the Materials Project - under Grant No. EDCBEE and supported
by Center for Next Generation Materials by Design, an Energy Frontier Research
Center funded by DOE, Office of Science, BES. Work at Lawrence Berkeley
National Laboratory was supported by the Office of Science of the U.S.
Department of Energy under Contract No. DEAC02-05CH11231. We thank the National
Energy Research Scientific Computing Center for providing invaluable computing
resources. Finally, we would like to thank all the users of the Materials
Project for their support and feedback in improving the project.

\bibliographystyle{wileyj}
\bibliography{gce15}

\newpage

\end{document}